\documentclass[12pt]{article}
\usepackage{epsfig}
\usepackage{graphicx,floatflt,amssymb,epsf}
\usepackage{graphics}
\global\parskip 6pt

\normalsize

\let\oldtheequation=\theequation
\def\doteqs#1{\setcounter{equation}{0}
            \def\theequation{{#1}.\oldtheequation}}

\newcounter{sxn}
\def\sx#1{\addtocounter{sxn}{1} \bigskip\medskip \goodbreak
\noindent{\large\bf 
{\thesxn.~~#1}} \nobreak \medskip}
\def\sxn#1{\sx{#1} \doteqs{\thesxn}}

\newcounter{axn}

\def\br{\begin{thebibliography}{abc}}
\def\er{\end{thebibliography}}

\def\be{\begin{equation}}

\def\ee{\end{equation}}

\def\bea{\begin{eqnarray}}

\def\eea{\end{eqnarray}}

\begin{document}
\begin{flushright}
\hfill{SINP-TNP/07-21}\\
\end{flushright}
\vspace*{1cm}
\thispagestyle{empty}
\centerline{\large\bf Universal near-horizon conformal structure and black hole entropy}
\bigskip
\begin{center}
Sayan K. Chakrabarti\footnote{Email: sayan.chakrabarti@saha.ac.in} $\&$ 
Kumar S. Gupta \footnote{Email: kumars.gupta@saha.ac.in}\\
\vspace*{.2cm}
{\em Theory Division\\ Saha Institute of Nuclear Physics\\
1/AF Bidhannagar, Calcutta - 700064, India}\\
\vspace*{.5cm}
Siddhartha Sen\footnote{Email: sen@maths.ucd.ie}\\
\vspace*{.2cm}
{\em School of Mathematical Sciences}\\
{\em UCD, Belfield, Dublin 4, Ireland}\\ 
\vspace*{.1cm}
{\em and}\\
\vspace*{.1cm}
{\em Department of Theoretical Physics}\\
{\em Indian Association for the Cultivation of Science}\\
{\em Calcutta - 700032, India}\\
\end{center}
\vskip.25cm

\begin{abstract}
  
  It is shown that a massless scalar probe reveals a universal
  near-horizon conformal structure for a wide class of black holes,
  including the BTZ. The central charge of the corresponding Virasoro
  algebra contains information about the black hole. With a suitable
  quantization condition on the central charge, the CFT associated
  with the black hole in our approach is consistent with the recent
  observation of Witten, where the dual theory for the BTZ in the AdS/CFT
  framework has been identified with the construction of Frenkel,
  Lepowsky and Meurman. This CFT admits the Fischer-Griess monster
  group as its symmetry. The logarithm of the dimension of a specific
  representation of the monster group has been identified by Witten as
  the entropy of the BTZ black hole. Our algebraic approach shows that
  a wide class of black holes share the same near-horizon conformal
  structure as that for the BTZ. With a suitable quantization
  condition, the CFT's for all these black holes in our formalism can be 
identified with the FLM model, although not through the AdS/CFT correspondence. 
The corresponding entropy for the BTZ provides a lower bound for the entropy of this entire class of black holes.

\end{abstract}

\vspace*{.3cm}
\begin{center}
August 2007
\end{center}
\vspace*{0.5cm}

PACS : 04.70.Dy \\
\newpage

\sxn{Introduction}

Conformal field theory plays a central role in the analysis of various
quantum aspects of black holes \cite{strom, carlipcft, solo}.  In particular,
for spaces with negative cosmological constant, the AdS/CFT duality
\cite{maldacena1, adscft} has led to remarkable insights in black hole physics.
This duality, which encodes the holographic principle \cite{gthooft,
  susskind, se} for AdS spaces,
provides a successful derivations of black hole entropy, consistent
with the Bekenstein-Hawking expression.

Recently, Witten \cite{witten} has used the AdS/CFT duality to obtain
the entropy of a quantum 2+ 1 dimensional BTZ black hole \cite{btz,
  btz2}. The BTZ black hole is associated with a Virasoro algebra
whose classical central charge, obtained using the method of Brown and
Henneaux \cite{brown}, is a continuous quantity. It has however been
argued that in the dual description the central charge $c$ is not
continuous, but is quantized as $24k$, where $k$ is a positive
integer. This is consistent with the Chern-Simons description
\cite{town, witten3} of BTZ
for a suitable choice of the gauge group. For $k=1$, Witten
\cite{witten} has identified the corresponding dual theory with the
$c=24$ holomorphic CFT obtained by Frenkel, Lepowsky and Meurman (FLM)
\cite{flm}. The FLM CFT describes only gravitational degrees of freedom and 
is not associated with any gauge fields. In addition, the $c=24$ FLM CFT is conjectured to be unique \cite{tuite} and admits a symmetry 
given by the Fischer-Griess monster group. Using this
symmetry, Witten calculated the entropy of the quantum BTZ black hole
as the logarithm of the dimension of a specific representation of the
monster group relevant for the FLM CFT. The result obtained is in
reasonable agreement with the semiclassical value for the
Bekenstein-Hawking entropy.

Another approach which has been very useful in black hole entropy
analysis is based on the existence of a CFT in the near-horizon region
of black holes \cite{carlipcft, kang, cvitan, camb}. The
analysis here is also based on Brown and Henneaux's approach
\cite{brown}, suitably adapted to the near-horizon region.  This
approach has led to the derivation of entropy for BTZ black hole as well
as for Schwarzschild black holes in arbitrary dimensions and the
results obtained are consistent with the Bekenstein-Hawking formula.

In a  previous paper we have shown that the near-horizon conformal structure can also be revealed by probing the black hole geometry with a 
massless scalar probe \cite{gupta}. Using the zero mode of a massless
scalar field, this approach was shown to be consistent with the
existence of a Virasoro algebra in the near-horizon region of 3+1
dimensional Schwarzschild black hole. Our approach also predicted the
logarithmic correction to the Bekenstein-Hawking entropy \cite{sen}. Similar 
scalar field probe can also be used to analyze particle production in black hole backgrounds \cite{paddy}.

In this paper we shall use the formalism developed in our earlier 
work \cite{gupta} to
analyze the near-horizon geometry of a large class of black holes,
including that for the BTZ. Unlike \cite{gupta}, here we use an
arbitrary frequency mode of a massless scalar field to probe the
near-horizon region.  The central charge of the near-horizon CFT is
obtained as a continuous variable. It is determined by the coefficient 
of the inverse square term in the near-horizon operator, which depends  
on the black hole parameters and the frequency of the probe. In order that our near-horizon CFT admits a central charge $c=24$ requires a quantization 
condition. This condition fixes the scalar field frequency in 
terms of the black hole parameters. Witten has suggested that the AdS/CFT dual of the BTZ black hole is the $c=24$ conformal field theory of FLM, which describes only gravitational degrees of freedom and is conjectured to be unique. In our analysis there is no AdS/CFT correspondence, but we find the possibility of a $c=24$ CFT for the BTZ. The uniqueness conjecture then implies that the $c=24$ CFT obtained in our analysis should also be identified with the FLM construction. In addition, we explicitly show that a large class of black holes
contain the same near-horizon conformal structure as that for the BTZ, all of which admit a $c=24$ CFT in the near-horizon region. From the uniqueness conjecture it follows that all these CFT's should also be identified with the FLM construction. The $c=24$ FLM CFT has the largest degeneracy of all allowed finite CFT's and hence provides a bound on the black hole entropy. The emergence of this
possibility in various backgrounds considered here is consistent with the fact that the BTZ black hole appears in the near-horizon region of many black holes \cite{maldacena, sen3, sfetsos, larsen, satoh, peet, gupta1, ashoke}.

This paper is organized as follows. In Section 2, we study the
dynamics of the scalar field probe in the near-horizon region for the
BTZ black hole. In Section 3, this analysis is generalized to include
a wide class of black holes. Different black hole near horizon
structures and the near horizon KG operator including the Gauss-Bonnet
case obtained from string derived gravity is discussed in the
Appendix. It is shown that the near-horizon dynamics of the scalar
probe is governed by an operator which has a universal form. In
Section 4, the algebraic properties of the near-horizon Klein-Gordon
operator is studied which reveals the existence of an underlying
conformal algebra consisting of a semi-direct product of the Virasoro
algebra and the algebra of the shift operators.  In Section 5, we
study the representations of this algebra and obtain a universal form
of the central charge. Section 6 discusses the connection of the
central charge to the black hole entropy in the context of the FLM CFT
obtained by Witten. Section 7 concludes the paper with some
discussions.

\sxn{Scalar field probe of the near-horizon geometry of BTZ black hole}

In 2+1 dimensions, a  BTZ black hole of mass $M$ and spin $J$ is given by the metric \cite{btz,btz2}
\bea
ds^2=-\left(-M+\frac{r^2}{l^2}+\frac{J^2}{4r^2}\right)dt^2+\left(-M+\frac{r^2}{l^2}+\frac{J^2}{4r^2}\right)^{-1}dr^2+r^2\left(d\phi-\frac{J}{2r^2}dt\right)^2,
\eea
where $r$ and $\phi$ denote the radial and angular coordinates on the plane and $t$ denotes the time.
This metric satisfies vacuum Einstein equations in $2+1$ dimensions,
with a negative cosmological constant $\Lambda=-1/l^2$.  The outer and inner horizons denoted by $r_{\pm}$ respectively are given by
\be
r_{\pm}^2=\frac{Ml^2}{2}\left(1\pm\sqrt{1-\frac{J^2}{M^2l^2}}\right).
\ee

In the present approach, a massless scalar field is used as a probe of the
near-horizon geometry.  The Klein-Gordon (KG) equation for a massless
scalar field $\psi$ in a general space-time with a metric
$g_{\mu\nu}$, where $\mu,\nu$ run over the space-time indices, is
given by
\begin{equation}
\frac{1}{\sqrt{-g}}\partial_{\mu}(\sqrt{-g}g^{\mu\nu}\partial_{\nu}\psi)=0.
\label{KG}
\end{equation}
For the BTZ metric (2.1), using the ansatz $\psi(r,t,\phi)=R(r)e^{-i\omega
  t+ik\phi}$, one gets the radial equation as \cite{sen1} 
\bea
\partial_r^2R(r)+\left(-\frac{1}{r}+\frac{2r}{r^2-r_-^2}+\frac{2r}{r^2-r_+^2}\right)\partial_rR(r)+N^{-4}\left(\omega^2-\frac{J\omega
  k}{r^2}+\frac{Bk^2}{r^2}\right)R(r)=0,
\label{KGBTZ}
\eea 
where $N^2=\frac{1}{l^2r^2}(r^2-r_-^2)(r^2-r_+^2)$ and
$B=M-\frac{r^2}{l^2}$. The periodicity of the coordinate $\phi$ in the
BTZ construction leads to the quantization condition $k\in
\mathcal{\bf Z}$.

Our main interest is to probe the region near the outer horizon of
this black hole. To this end, we define a near-horizon coordinate $x
\in [0, \infty)$ as 
\be 
x=r-r_+.  
\ee 
In the near-horizon region, the
function $N^2$ takes the form 
\be 
N^2(x)=Ax, ~~~A =
\frac{2}{l^2r_+}(r_+^2-r_-^2).  
\ee 
We now define a new radial
wave-function $\chi(r) \equiv \sqrt{x}R(r)$, in terms of which the
near-horizon form of the KG equation (\ref{KGBTZ}) is given by 
\be 
H\chi(r) \equiv \left [ -\frac{d^2}{dx^2}+\frac{a}{x^2} \right ]
\chi(r), \label{btzop} 
\ee 
where 
\be 
a =-\left[\frac{1}{4}+\frac{{\tilde{\omega}}^2}{A^2}\right],~~~
{\tilde{\omega}}^2 =\omega^2-\frac{J\omega
  k}{r_+^2}+\frac{Mk^2}{r_+^2}-\frac{k^2}{l^2}. \label{P} 
\ee 
In (\ref{btzop}), $H$ denotes the near-horizon KG operator for the BTZ
black hole. The parameter $a$ contains the information specific to the geometry.  Below we shall show that the near-horizon KG operator for
a large class of black holes has the same form as in (\ref{btzop}) with
different values of the parameter $a$.

\sxn{Universal near-horizon geometry for a large class of black holes}

Consider a metric in $D$ space-time dimensions given by
\be
ds^2=-f(r)dt^2+\frac{dr^2}{f(r)}+r^2d\Omega_{D-2}^2, \label{metric}
\ee 
where $f(r)$ is a function 
of the radial variable $r$ and $d\Omega_{D-2}^2$ is the metric on unit $S^{D-2}$. The form of the functions $f(r)$ will depend on the choice of the specific black hole.

As before, we wish to probe the near-horizon geometry of the black
hole using a massless scalar field. To that end, we again consider the
dynamics of a scalar field $\psi$ in the above general background.
Using the ansatz $\psi(t,r,\Omega)=e^{-i\omega t}R(r)Y_{lm}(\Omega)$,
the KG equation for the general background (3.1) can be written as 
\bea
\frac{d^2R(r)}{dr^2}+\frac{(f(r)r^{D-2})^{\prime}}{f(r)r^{D-2}}\frac{dR(r)}{dr}+\frac{\omega^2R(r)}{f^2(r)}-\frac{1}{f(r)}\left(\frac{l(l+D-3)}{r^2}\right)R(r)=0,
\label{KG1}
\eea  
where the prime denotes the derivative with respect to the radial
coordinate.

As in the case for BTZ, we now define a near-horizon coordinate $x\equiv r-r_h$, where $r_h$ denotes the event horizon and $x \in [0,\infty)$. We shall show below that the near-horizon form of the function $f$ for a large class of black holes is given by
\be
f(x)\sim{A}x[1+\mathcal{O}(x)],
\ee
where $A$ is a
constant which depends on parameters defining the black hole geometry. 
For the moment we proceed with the form of $f$ given in (3.3).
In terms of a new field $\chi \equiv \sqrt{x}R(r)$, the KG equation in the near-horizon region takes the form 
\be
H \chi = \left [ -\frac{d^2}{dx^2} + \frac{a}{x^2} \right ] \chi ,
\label{KGnear}
\ee
where
\be 
a = - \left[\frac{1}{4} + \frac{\omega^2}{{A}^2} \right].\label{otherop}
\ee

The above analysis thus shows that for the entire class of metrics (3.1)
satisfying the condition (3.3), the near-horizon KG operator $H$ for a massless 
scalar has a universal form given by
\bea 
H=-\frac{d^2}{dx^2}+\frac{a}{x^2}.
\label{KGop1}
\eea
The structure of $H$ here is the same as that for the BTZ black hole as 
given by (\ref{btzop}). The constant $a$ depends on the frequency of the 
probe and on the geometric details and distinguishes between the various 
black holes. For real and nonzero frequency of the scalar probe the constant 
$a$ is real and satisfies the condition $a < -\frac{1}{4}$.

The above analysis shows that for a very large class of black
holes, the near-horizon KG operator for a massless scalar has a
universal form. This result is somewhat surprising as the black holes
have quite different geometric properties. We have discussed different
black hole metrics in detail in the appendix and it can be seen from
the discussion in the appendix that apart from asymptotically flat
backgrounds, our analysis includes black holes with both signs of the
cosmological constant. It also includes the Gauss-Bonnet case which is
obtained from string derived gravity going beyond the usual
Einstein-Hilbert action. In a later section we shall comment on the
possible common link between the near-horizon structures of these
varied class of black holes with that for BTZ. Before that we shall
study the algebraic properties of the near-horizon KG operator.

\sxn{Algebraic properties of the near-horizon KG operator}

We have seen from the previous section that for a large class of black
holes, the KG operator in the near-horizon region has a universal form
given by (3.6). In this section we shall use algebraic techniques to
study the properties of $H$.

Following our earlier work \cite{gupta,rajeev}, the operator $H$ can
be decomposed as
\bea
H=C_+C_-,~~~\mbox{where}~~~ C_{\pm}=\pm\frac{d}{dx}+\frac{b}{x},
\eea
and 
\bea
b=\frac{1}{2}\pm\frac{\sqrt{1+4a}}{2}. \label{b}
\eea
For the present cases of interest, $a<-\frac{1}{4}$, $b$ is complex and
$C_+$ and $C_-$ are not formal adjoints of each other. Following
\cite{gupta} we define the operators
\bea
L_n&=&-x^{n+1}\frac{d}{dx}, ~~~~n\in\mathcal{\bf Z},\label{l}\\
P_m&=&\frac{1}{x^m}, ~~~~m\in\mathcal{\bf Z}\label{p}
\eea
Using Equation (\ref{l}) and (\ref{p}), the operators $C_{\pm}$ and
$H$ can be written as
\bea
C_{\pm}&=&\mp L_{-1}+bP_1,\\
H&=&(-L_{-1}+bP_1)(L_{-1}+bP_1).
\eea
%
%\newpage
The operators $L_m$, $P_m$ and $H$ satisfy the commutation relations
\bea {}
[ P_{m}, P_{n} ] &=& 0,\label{pp}\\ {}
[ L_{m}, P_{n} ] &=& n P_{n - m},\label{lp}\\ {}
[ L_{m}, L_{n} ] &=& ( m - n) L_{m + n} + \frac{c}{12}
(m^{3} - m)\delta_{m+n, 0},\label{central}\\ {}
[P_m, H] &=& m (m+1) P_{m + 2} + 2 m L_{-m -2},\label{ph} \\{}
[L_m, H] &=& 2b(b-1)P_{2-m} - (m+1)(L_{-1}L_{m-1} +
L_{m-1}L_{-1}).\label{lh} 
\eea
Note that Equation (\ref{central}) describes a Virasoro algebra with
central charge $c$, while the algebra of the generators defined in
Equation (\ref{l}) would lead to $[L_m,L_n]=(m-n)L_{m+n}$. However,
this algebra admits a non-trivial central extension $c$ and in any of
its unitary irreducible highest weight representation $c \neq 0$
\cite{goddard}. That is why the central charge has been explicitly
included in (\ref{central}).

Equations (\ref{pp}-\ref{central}) describe the semidirect product of
the Virasoro algebra with an Abelian algebra satisfied by the shift
operators ${P_m}$ \cite{raina}. This semidirect product
algebra will be denoted by $\mathcal{M}$ in the rest of this 
paper. Note that $L_{- 1}$ and $P_1$ are the only  generators that appear in $H$.
Starting with these two generators, and using (\ref{ph}) and (\ref{lh}),
we see that the only operators which appear are the Virasoro generators
with negative index (except $L_{-2}$) and the shift
generators with positive index.
Thus, $L_m$ with $m \geq 0$ and $P_m$
with $m \leq 0$ do not appear in the above expressions. In the next
section, we will discuss how these quantities are generated.

We also note that the operator $H$ is not an element of $\mathcal{M}$
but belongs to the corresponding enveloping algebra.  This is due to
the fact that the right hand side of Equation (\ref{lh}) contains
product of Virasoro generators. While such products are not elements
of the algebra, they do belong to the corresponding enveloping
algebra.

\sxn{Representation theory}

The representation theory of $\mathcal{M}$ is well known in the literature \cite{raina}. Below we briefly recall certain aspects of this theory relevant for our analysis.

Consider the space $V_{\alpha,\beta}$ of densities containing
elements of the form $P(x)x^{\alpha}(dx)^{\beta}$. Here, $\alpha$,
$\beta$ are complex numbers and $P(x)$ is an arbitrary polynomial in
$x$ and $x^{-1}$. Note that $x$ is now treated as a complex variable
and the algebra $\mathcal{M}$ remains unchanged even when $x$ is
complex. $V_{\alpha, \beta}$ carries a representation of the algebra
$\mathcal{M}$. The space $V_{\alpha,\beta}$ is spanned by a set of
basis vectors $\omega_{m}=x^{m+\alpha}(dx)^{\beta}$, where
$m\in\mathcal{\bf Z}$. It can be shown that the Virasoro
generators and the shift operators have the following action on the
basis vectors $\omega_{m}$ \cite{raina}
\bea
P_n (\omega_m) &=& \omega_{m-n},\label{pw}\\
L_n (\omega_m) &=& - (m + \alpha + \beta + n \beta) \omega_{n + m}.\label{lw}
\eea 
The representation $V_{\alpha, \beta}$ is reducible if $\alpha \in
{\mathbf Z}$ and if $\beta = 0$ or $1$; otherwise it is irreducible.

The requirement of unitarity of the representation $V_{\alpha, \beta}$
leads to several important consequences. In any unitary representation
of ${\cal {M}}$, the Virasoro generators must satisfy the condition
${L_{- m}^{\dagger}} = L_m$. In the previous section, we saw that
$L_{-2}$ and $L_m$ for $m \geq 0$ did not appear in the algebraic
structure generated by the basic operators appearing in the
factorization of $H$.  However, the requirement of a unitary
representation now leads to the inclusion of $L_m$ for $m > 0$. The
remaining generators now appear through appropriate commutators, thus
completing the algebra ${\cal M}$.

Unitarity also constrains the parameters $\alpha$ and $\beta$, which must
satisfy the conditions 
\bea
\beta + \bar{\beta} &=& 1,\label{bbbar}\\
\alpha + \beta &=& {\bar{\alpha}} + {\bar{\beta}},\label{ababarbbar}
\eea
where $\bar{\alpha}$ denotes the complex conjugate of $\alpha$. The
central charge $c$ in the representation $V_{\alpha,\beta}$ is then
given by \cite{raina}
\bea
c(\beta)=-12\beta^2+12\beta-2. \label{ccharge}
\eea

Next we study the quantum properties of the Klein-Gordon operator $H$
in the near horizon region of the black holes. In particular, we shall
analyze the eigenvalue equation
\bea
H |\Psi\rangle=E |\Psi\rangle, \label{ee}
\eea
using the representations of the algebra $\mathcal{M}$ and we shall 
find the labels $\alpha$ and $\beta$ of the representation of
$\mathcal{M}$ in terms of the black hole parameters.

We choose an ansatz for the wave function $|\Psi\rangle$ given by
$|\Psi\rangle = \sum^{\infty}_{n=0} c_n \omega_n$. The indicial
equation arising out from the substitution of the above ansatz in the
eigenvalue equation (\ref{ee}) is given by
\bea
\alpha=b,~~ \mbox{or}~~ (1-b) \label{indicial}
\eea
As discussed before, for nonzero real frequency of the probe, the constant $a$ is real and satisfies the condition $a<-\frac{1}{4}$.
Thus, in the general case, we can use the parametrization
\bea
a=-\frac{1}{4}-\mu^2 \label{a}.
\eea
The information about the black hole geometry is contained in 
the parameter $\mu \in R$.
Using (\ref{a}) in (\ref{b}), we obtain
\bea
b=\frac{1}{2}\pm i\mu.
\eea
Equations (\ref{indicial}) and (5.9) determine the label $\alpha$ in the representation of  $\mathcal{M}$, whose allowed values are given by 
\bea
\alpha=\frac{1}{2}\pm i\mu,~~ \mbox{or}~~\alpha=-\frac{1}{2}\mp i\mu.
\eea  
Using (5.3) and (5.4), the other parameter $\beta$ in the representation of 
$\mathcal{M}$ is determined in terms of $\alpha$ as
\be
\beta = \frac{1}{2} - {\rm Im}~\alpha.
\ee

The central charge $c$ in any representation of $\mathcal{M}$ depends
only on the value of $\beta$. Using (5.5), (5.10) and (5.11), the
central charge is obtained as 
\be 
c = 1 + 12 \mu^2.  
\ee 
This expression of the central charge has two contributions. The first
part is a constant 1, which is independent of the black hole
parameters. The second part depends on $\mu$, which contains
information about the black hole parameters. In our framework, we
associate the quantity $ c_{bh} \equiv 12 \mu^2 $ with the
contribution to the central charge due to the black hole, as detected by 
the scalar field probe. In the subsequent analysis we shall focus on 
$c_{bh}$ only.

\sxn{Central charge and black hole entropy}

The relation between CFT and black hole entropy has been extensively
discussed in the literature \cite{strom, carlipcft, solo}.  
%Recently
%Witten has proposed to calculate the entropy of the BTZ black hole
%from the dual conformal field theory \cite{witten}, within the general
%framework of AdS/CFT correspondence.  
In the Chern-Simons approach to
2+1 gravity with a negative cosmological constant \cite{town, witten3}, the method of Brown
and Henneaux \cite{brown} predicts a continuously varying central charge. This is a
purely classical result where nothing in this analysis puts any
constraint on the central charge. However, motivated by the AdS/CFT
correspondence, Witten \cite{witten} argued that the central charge cannot
vary continuously, but must be quantized. In the Chern-Simons approach to
BTZ, this assumption leads to a central charge given by $c=24k$, where $k$ is
the coefficient in front of the Chern-Simons action. The constant $k$, for a
suitable choice of the gauge group, is quantized as a
positive integer. Assuming $k=1$ for the moment, the central charge
has the value $c=24$. Of all the holomorphic CFT's associated with
$c=24$, there is only one which corresponds to pure gravity. Such a
CFT model was explicitly constructed by Frenkel, Lepowsky and Meurman
(FLM)\cite{flm}, who also conjectured that it would be unique. This
model admits a huge symmetry, given by the Fischer-Griess monster
group. Using the FLM representation and exploiting its uniqueness,
Witten obtained the entropy of the quantum BTZ black hole which agrees
reasonably well with the Bekenstein-Hawking entropy.

In our formalism, the central charge associated with the black hole
degrees of freedom is given by $c_{bh} = 12 \mu^2$. From the pure
classical gravity point of view, there is no constraint on this central
charge.  However, motivated by the same logic of Witten, we also demand
quantization of the central charge for the BTZ. In particular, we impose the
condition $\mu^2 = 2 n$, where $n$ is a positive integer. 
Then, for $n=1$, we have a CFT with central charge equal to 24. The FLM
construction is conjectured to give a unique CFT for pure gravity with
$c=24$. Thus, the above quantization condition together with the uniqueness 
conjecture for the FLM construction implies that the near-horizon CFT for 
the BTZ black hole in our approach as well is given by the FLM CFT. Consequently, the 
entropy calculation of Witten for the BTZ black hole based on the monster symmetry of 
the FLM CFT would apply in our framework too. 
Let us note that our argument is valid only with the quantization condition 
$\mu^2 = 2n$. The quantity $\mu^2$ depends on the scalar field frequency and
also on the geometric parameters defining the BTZ black hole. This
condition implies that the scalar field frequency 
is quantized and that the scale of the frequency is set by the geometric 
parameters of the BTZ. These consequences are completely 
natural and such a quantization of the frequency is also in qualitative
agreement with the brick wall approach to black hole entropy 
\cite{thooft, guptabw}.

The present analysis also indicates that the near-horizon conformal
structure for the BTZ is shared by a wide class of black holes. In
particular, the near-horizon KG operator (3.6) and the central charge
(5.12) have the same universal form for all the black holes considered
here. For this larger class, we can again identify $12\mu^2$ in the
central charge as the contribution associated to the black hole
geometry, as this part depends explicitly on the black hole
parameters. In string theory it is well known that a large class of black 
holes, including the Schwarzschild case, contains a BTZ factor as a part of their
near-horizon geometry \cite{adscft, maldacena, sen3,
sfetsos, larsen, satoh, peet, gupta1, ashoke}.  The result that a 
large class of black holes share a universal near-horizon conformal structure 
identical to that of the BTZ is thus consistent with the observations from string theory. 

The existence of a BTZ factor in 
the near-horizon region for a large class of black holes has been very useful 
in calculating the black hole entropy in string theory \cite{peet, ashoke}. Motivated by 
this observation, we can conjecture that the central charge for the entire class of black holes that share the same near-horizon conformal structure as that for the BTZ should also be subject to a similar quantization. Imposition of the same quantization condition as in BTZ
leads to $c_{bh} = 24$ in each of these cases. Moreover, the scalar
field frequency for each background would now be quantized, where the
scales in the frequencies would be set by the respective black hole
parameters. Since we are dealing with pure gravity, the FLM
construction would provide the unique CFT in each of these cases,
leading to a universal form of entropy for the entire class. This
presents an apparent puzzle, as surely the Bekenstein-Hawking entropy
for a Schwarzschild black hole is not going to be the same as that for
BTZ. This can be resolved by noting that the near-horizon geometry for a general
black hole contains factors other than the BTZ, which can also contribute to
the entropy. Thus in our analysis, the BTZ entropy contribution provides a
lower bound to the entropy of this entire class of black holes.

\sxn{Conclusion}

In this paper we have analyzed the near-horizon conformal structure
for a large class of black holes, using a scalar field as a probe of
the geometry. The class of backgrounds considered here includes black
holes in various dimensions, with or without cosmological constant. We
have also considered Gauss-Bonnet black hole, which is obtained from
string derived gravity by going beyond the usual Einstein-Hilbert
action of general relativity.

The analysis presented here reveals certain universal characteristics
in the near-horizon conformal structure. In particular, the
near-horizon KG operator and the central charge of the associated
Virasoro algebra has the same universal form for the entire class of
black holes discussed here.  Our analysis is consistent with the
observation in string theory that a large class of black holes
contains a BTZ part as their near-horizon geometry. The BTZ entropy
has recently been obtained by Witten using the FLM construction. Although 
we do not have an AdS/CFT correspondence, our
analysis for the BTZ black hole is consistent with Witten's
observations after a suitable quantization condition is imposed.
Furthermore, assuming that such a quantization can be imposed for the
other cases considered here, it appears plausible that the FLM CFT
provides a universal contribution to the entropy for a large class of
black holes. The actual entropy would depend on other factors as well,
but the universal part in our approach is determined by the FLM model, which
provides a lower bound to the entropy of this wide class of black holes.
For any given background, the assumed condition on the central charge also 
leads to the quantization of the scalar field frequency, where the scale is 
set by the parameters defining the respective black hole.

In this paper, although black holes with positive cosmological
constants have been analyzed, the physical interpretation in such
cases are subject to usual ambiguities associated with de-Sitter
spaces \cite{witten2, susskind2}. It would be interesting to see if some of those ambiguities
can be resolved in the framework of the near-horizon CFT discussed here. 
We have also not discussed the
extremal black holes, which would also be of interest.

%\newpage

\sxn{Appendix}

In this appendix we shall give examples of different black hole
metrics for which the function $f$ has the form as given in (3.3).  We
shall first discuss the examples where the cosmological constant
vanishes

\noindent
{\bf {I. Asymptotically flat black hole geometries}} \\
 
\noindent{\bf Schwarzschild black hole:}\\

\noindent
The metric function $f(r)$ is given by
\bea
f(r)=1-\frac{2M}{r^{D-3}}.
\eea
The event horizon, determined by $f(r)=0$, is located at 
\bea
r_h=(2M)^{\frac{1}{D-3}},
\eea
so that $f(r)\geq 0$ for $r\geq r_h$ and the parameter $M$ is always
positive. Now, the form of the metric near the horizon is given by
\bea
f(x)=Ax,~~\mbox{where}~~A=\frac{2M(D-3)}{r_h^{D-2}}.
\eea
Here $A$ is a constant for a particular $M$.\\ 

\noindent
{\bf Reissner Nordstr\"{o}m black hole:}\\

\noindent
The function $f(r)$ has the following form
\bea
f(r)=1-\frac{2M}{r^{D-3}}+\frac{Q^2}{r^{2D-6}},
\eea
and the event horizon corresponds to 
\bea
r_h^{\pm}=\left(M\pm\sqrt{M^2+Q^2}\right)^{\frac{1}{D-3}}.
\eea
Here, we are mainly interested in the outer horizon, $r_h^+$, since
all our calculations will eventually involve the near-horizon
coordinate and here `near-horizon' implies near the outer horizon. In
this case $f(r)\geq 0$ for $r\geq r_h^+$. This essentially leads to
the constraint $Q^2\leq M^2$. Now, the form of the metric near the
horizon is given by
\bea
f(x)=Ax,~~\mbox{where}~~A=\left[\frac{2M(D-3)}{{r_h^+}^{D-2}}-\frac{Q^2(2D-6)}{{r_h^+}^{2D-5}}\right].
\eea
\\

\noindent{\bf Uncharged Gauss-Bonnet black hole:}\\

\noindent
In $D$ space-time dimension $(D\geq 5)$, the metric for spherically
symmetric asymptotically flat Gauss-Bonnet black hole of mass
$M$ is given by Eqn.  (\ref{metric}), where $f(r)$ has the form
\cite{deser}
\be
f(r) = 1 + \frac{r^2}{2\alpha} -
\frac{r^2}{2\alpha}\sqrt{1+\frac{8\alpha M}{r^{D-1}}}, \label{fr}
\ee
where
\be
\alpha=16\pi G_D (D-3)(D-4)\alpha'.
\ee
For $\alpha'>0$, this black hole admits only a single horizon
\cite{deser}. The horizon $r=r_h$ is determined by the real positive
solution of the equation
\be
r_h^{D-3}+\alpha r_h^{D-5}=2M. \label{horizon}
\ee
In terms of the near-horizon coordinate the function $f(r)$
becomes 
\bea
f(x)=Ax,~~\mbox{where}~~A=\left[\frac{r_h^2(D-3)+\alpha(D-5)}{r_h(r_h^2+2\alpha)}\right].
\eea
\\\\

\noindent
{\bf {II. Asymptotically non-flat black hole geometries: }}\\

\noindent{\bf Schwarzschild-AdS black hole:}\\

\noindent
The metric in this case is given by
\bea
f(r)=1-\frac{2M}{r^{D-3}}-\Lambda r^2,
\eea
where $\Lambda<0$ and $M>0$. There is an event horizon defined by
$f(r)=0$ and given by the unique, real and positive root of
\bea
|\Lambda|r^{D-1}+r^{D-3}-2M=0.
\eea
There is no cosmological horizon here. The form of the metric
near the horizon is given by
\bea
f(x)=Ax,~~\mbox{where}~~A=\left[\frac{2M(D-3)}{r_h^{D-2}}-2r_h \Lambda \right].
\eea
\\
\noindent{\bf Reissner Nordstr\"{o}m-AdS black hole:}\\

\noindent
The function $f(r)$ has the following form
\bea
f(r)=1-\frac{2M}{r^{D-3}}+\frac{Q^2}{r^{2D-6}}-\Lambda r^2, ~~~ \Lambda<0.
\eea
The event horizons are given by the solutions of the equation
\bea
|\Lambda|r^{2D-4}+r^{2D-6}-2Mr^{D-3}+Q^2=0,
\eea
and there is no cosmological horizon. This equation has two real
positive zeroes which gives the inner and outer horizons of the black hole. But, $f(r)\geq 0$
for $r\geq r_h^+$. Here also we are interested in the outer
horizon. Here $Q^2<M^2$ and $\Lambda <0$. The function $f(r)$ in the
near horizon coordinate has the form
\bea
f(x)=Ax,~~\mbox{where}~~A=\left[\frac{2M(D-3)}{{r_h^+}^{D-2}}-\frac{Q^2(2D-6)}{{r_h^+}^{2D-5}}-2r_h^+ \Lambda \right].
\eea
\\
\noindent{\bf Schwarzschild-dS black hole:}\\

\noindent
The metric has the same form as of Schwarzschild-AdS black hole,
except the fact that the cosmological constant $\Lambda>0$ here. There
is both an event horizon $r_h$ as well as a cosmological horizon
$r_c$ which are given by the real positive roots of 
\bea
\Lambda r^{D-1} - r^{D-3} + 2M=0.
\eea
This equation has two real positive zeroes corresponding to the event
and the cosmological horizon. The function $f(r)\geq 0$ for $r_h \leq
r \leq r_c$. We are interested in the region outside the event horizon
only. The form of the constant $A$ is given by
$A=\left[\frac{2M(D-3)}{r_h^{D-2}}-2r_h \Lambda \right]$, i.e. it is
same as that of Schwarzschild AdS black hole except here
the cosmological constant is positive.\\

\noindent{\bf Reissner Nordstr\"{o}m-dS black hole:}\\

\noindent
The metric here also looks like the same as of RN-AdS case except the
cosmological constant being positive here. There are two event
horizons now and a cosmological horizon given by the zeroes of 
\bea
\Lambda r^{2D-4} - r^{2D-6} + 2Mr^{D-3} - Q^2=0.
\eea

This equation now has three real positive roots corresponding to inner
and outer event horizons and a cosmological horizon. Here $f(r)\geq 0$
for $r_h^+ \leq r \leq r_c$. Here also the structure of $A$ is same as
that of the RN-AdS case but with positive cosmological constant.

\bibliographystyle{unsrt}

\end{document}